\documentclass[12pt,a4paper]{article}
\usepackage[truedimen,margin=30mm]{geometry}

\usepackage[dvipdfmx]{graphicx}
\usepackage{mathrsfs}
\usepackage{amssymb}
\usepackage{amsmath}
\usepackage{ascmac}
\usepackage{amsthm}
\usepackage{natbib}
\usepackage{setspace}
\usepackage{color}
\usepackage{times}   % font (times new roman)

\usepackage{titlesec}
\titleformat*{\section}{\large\bfseries}
\titleformat*{\subsection}{\it}

\def\ep{{\varepsilon}}

\title{{\bf Hierarchical Regression Discontinuity Design: Pursuing Subgroup Treatment Effects }}

\date{}

\begin{document}

\maketitle
\doublespacing

\vspace{-1.7cm}
\begin{center}
{\large Shonosuke Sugasawa$^{1\ast}$, Takuya Ishihara$^2$ and Daisuke Kurisu$^3$}

\medskip

\medskip
\noindent
$^1$Faculty of Economics, Keio University\\
$^2$Graduate School of Economics and Management, Tohoku University\\
$^3$Center for Spatial Information Science, The University of Tokyo\\
$^{\ast}$Corresponding (Email: sugasawa@econ.keio.ac.jp)

\bigskip
%\today
\end{center}

\vspace{0.3cm}
\begin{center}
{\bf \large Abstract}
\end{center}

Regression discontinuity design (RDD) is widely adopted for causal inference under intervention determined by a continuous variable. While one is interested in treatment effect heterogeneity by subgroups in many applications, RDD typically suffers from small subgroup-wise sample sizes, which makes the estimation results highly instable. To solve this issue, we introduce hierarchical RDD (HRDD), a hierarchical Bayes approach for pursuing treatment effect heterogeneity in RDD. A key feature of HRDD is to employ a pseudo-model based on a loss function to estimate subgroup-level parameters of treatment effects under RDD, and assign a hierarchical prior distribution to ``borrow strength" from other subgroups. The posterior computation can be easily done by a simple Gibbs sampling, and the optimal bandwidth can be automatically selected by the Hyv\"{a}rinen scores for unnormalized models. We demonstrate the proposed HRDD through simulation and real data analysis, and show that HRDD provides much more stable point and interval estimation than separately applying the standard RDD method to each subgroup. 
\vspace{-0cm}


\newpage
%%--------------------------------------------------------------------------%
%%     Introduction
%%--------------------------------------------------------------------------%
\section{Introduction}

Regression discontinuity design (RDD) is a representative method for causal inference under intervention determined by a continuous (running) variable and is widely adopted in a variety of areas such as economics, marketing, epidemiology, and clinical research.   
RDD was first introduced by \cite{thistlethwaite1960regression}, and then \cite{hahn2001identification} formalized the assumptions required for identification and proposed local linear estimators.
Later, \cite{imbens2008regression} and \cite{lee2010regression} surveyed the applied and theoretical literature on the RDD.
Further, \cite{imbens2012optimal} investigated the optimal bandwidth selection in terms of squared error loss, and \cite{calonico2014robust} proposed robust confidence interval estimators.

Although conventional RDD methods aim at estimation and inference on average treatment effects with adjusting potential effects of covariates other than the running variable, there is an increasing demand for pursuing treatment effect heterogeneity over subgroups determined by covariance. 
While a considerable number of methods for estimating heterogeneous treatment effects under randomized controlled trials and observational studies have been proposed \citep[e.g.][]{wager2018estimation,kunzel2019metalearners,hahn2020bayesian,sugasawa2023bayesian}, effective methods for pursuing treatment effect heterogeneity under RDD are scarce. 
One main reason is that in RDD the essential sample size to estimate the treatment effect can be much smaller than the total sample size since RDD fits local regression around the threshold value.
Therefore, this issue will be more severe when pursuing subgroup-wise treatment effects, and it could be unrealistic to conduct estimation and inference with reasonable accuracy.

To solve the aforementioned issue, we introduce hierarchical RDD (HRDD), a hierarchical Bayes approach for pursuing treatment effect heterogeneity in RDD.
A key feature of HRDD is to employ a pseudo-model based on a quadratic function to estimate subgroup-level parameters of treatment effects under RDD and assign a hierarchical prior distribution to ``borrow strength" from the information of other subgroups.
The estimation and inference on subgroup-wise treatment effects can be done through the posterior distributions, which can be easily simulated by a Gibbs sampling algorithm. 
A notable feature of HRDD can provide stable estimation and inference through hierarchical priors for subgroup treatment effects as well as subgroup-wise coefficients of local regression. 
As will be shown in our simulation study in Section~\ref{sec:sim}, HRDD can significantly improve the performance of separate RDD (that simply applies RDD for each subgroup). 
Furthermore, we apply HRDD to Colombia scholarship data with 106 subgroups determined by five categorical covariates and show that HRDD can produce stable estimation results of subgroup-wise treatment effects while addressing potential heterogeneity.
Moreover, HRDD for continuous responses can be extended to binary responses by changing the quadratic loss function for treatment effects to the negative log-likelihood of logistic models.  
Moreover, owing to the flexibility of the Bayesian framework, we can incorporate potential sparsity in the treatment effects by employing spike-and-slab priors \citep{ishwaran2005spike} for the treatment effect parameters. 
Further, we propose using the Hyv\"{a}rinen score \citep{hyvarinen2005estimation} to select the optimal bandwidth parameter within the MCMC iteration.

Regarding the existing methods for addressing potential heterogeneity in RDD, \cite{bartalotti2017regression} consider the designs with clustering structure and provide cluster-robust optimal bandwidth selectors in RDD, but they do not consider the heterogeneity of treatment effects and assume that treatment effects are the same across groups.
Further, \cite{becker2013absorptive} propose heterogeneous treatment effects estimators and quantify the heterogeneity of treatment effects of the EU’s main regional transfer program, and more recently \cite{keele2015geographic} and \cite{sawada2024local} investigate the heterogeneity of treatment effects in geographic and multivariate RDDs, respectively.
However, these existing methods for heterogeneous RDD do not incorporate a structure of ``borrowing strength" unlike HRDD, so that the performance could be highly affected by the subgroup-wise sample size. 
Regarding Bayesian approaches to RDD, \cite{chib2023nonparametric} proposed a nonparametric Bayesian method to estimate average treatment effects under RDD and is not concerned with subgroup heterogeneity of treatment effects.

This paper is organized as follows. 
In Section~\ref{sec:HRDD}, we introduce HRDD and detailed settings and implementation under both continuous and binary responses. 
In Section~\ref{sec:sim}, we demonstrate the performance of HRDD together with methods separately applying the conventional RDD to each subgroup. 
Section~\ref{sec:app} provides an application of HRDD to Colombia scholarship data and Section~\ref{sec:conc} gives concluding remarks. 
Details of posterior computation algorithms are deferred in the Appendix.

%--------------------------------------------------------------------------%
%     Method
%--------------------------------------------------------------------------%
\section{Hierarchical regression discontinuity design}\label{sec:HRDD}
\subsection{Hierarchical models}
Consider the standard setup of regression discontinuity design (RDD). 
Let $Y_{ig}$ be a study variable of interest of the $i$th sample in the $g$th subgroup, where $i=1,\ldots,n_g$ and $g=1,\ldots,G$.
Further, let $X_{ig}$ be a running variable and the treatment indicator is defined as $W_{ig}=I(X_{ig}\geq c)$ for some $c$.
Although we assume the same threshold value $c$ for all the groups for simplicity, it is straightforward to extend this setting to subgroup-wise threshold values $c_g$. 
The conditional average treatment effect at $X=c$ is defined as 
$$
\tau_g=\lim_{x\to c+0}E[Y_{ig}|X_{ig}=x]-\lim_{x\to c-0}E[Y_{ig}|X_{ig}=x], \ \ \ \ g=1,\ldots,G.
$$
For estimating $\tau_g$, a standard approach is fitting local linear regression to treatment and control regions. 
This is equivalent to estimating the weighted linear regression defined as the following loss function: 
\begin{equation}\label{loss}
L(\tau_g, \beta_g; h_g)=\frac12\sum_{i=1}^{n_g}K\left(\frac{|X_{ig}-c|}{h_g}\right)\left(Y_{ig}-\tau_gW_{ig}-Z_{ig}(c)^\top\beta_g\right)^2, \ \ \ \ g=1,\ldots,G,
\end{equation}
where $Z_{ig}(c)=(1, (X_{ig}-c)_{-}, (X_{ig}-c)_{+})$ and $K(\cdot)$ is a weighting function. 
It is also possible to use local polynomial regression with arbitrary order, so we let the dimension of $Z_{ig}(c)$ be $p$ in what follows. 
For example, if we use $q$th order polynomial, then $Z_{ig}(c)=(1, (X_{ig}-c)_{-}, (X_{ig}-c)_{+},\dots,(X_{ig}-c)_{-}^{q}, (X_{ig}-c)_{+}^{q})$, so that $p=2q+1$. 
Popular choices for the weighting function include the window function $K(x) = I({|x|\leq 1})$ or the triangular kernel $K(x) = (1-|x|)_{+}$.
In many applications, the subgroup-wise sample sizes $n_g$ tend to be small and the estimator of $\tau_g$ obtained by minimizing (\ref{loss}) can be unstable.

Consider a pseudo statistical model for $Y_g=(Y_{1g},\ldots,Y_{n_gg})$, defined as 
$$
p(Y_g | \tau_g, \beta_g, \omega )\propto \exp\{-\omega L(\tau_g, \beta_g; h_g)\},
$$
where $\omega$ is a universal scaling constant.
The use of a pseudo model to define posterior distributions is unknown as the general Bayesian method, and its decision theoretic justification is provided in \cite{bissiri2016general}. 
The above pseudo model can be seen as the power of the normal distribution, expressed as 
\begin{equation}\label{model1-plain}
p(Y_{ig} | \tau_g, \beta_g, \omega ) \propto  \phi(Y_{ig}; \tau_gW_{ig}+Z_{ig}^\top\beta_g, \omega^{-1})^{k_{ig}}, 
\end{equation}
independently for $i=1,\ldots,n_g$, where $k_{ig}=K(|X_{ig}-c|/h_g)$ and $Z_{ig}\equiv Z_{ig}(c)$.
To make the pseudo model (\ref{model1-plain}) against potential outliers, we introduce local parameters as follows: 
\begin{equation}\label{model1}
p(Y_{ig} | \tau_g, \beta_g, \omega, u_{ig} ) \propto  \phi(Y_{ig}; \tau_gW_{ig}+Z_{ig}^\top\beta_g, \omega^{-1}u_{ig}^{-1})^{k_{ig}}, 
\end{equation}
where the local parameter $u_{ig}$ follows the two-component mixture distribution as follows: 
$$
u_{ig}|(r_{ig}=0)\sim \delta_1, \ \ \ u_{ig}|(r_{ig}=1)\sim  {\rm Ga}(\nu, \nu), \ \ \ P(r_{ig}=1)=1-P(r_{ig}=0)=w.
$$
Here ${\rm Ga}(a, b)$ denotes a gamma distribution with shape parameter $a$ and rate parameter $b$, and $\delta_1$ is the one-point distribution on $u_{ig}=1$, and $\nu$ is a tuning parameter. 
Note that observations with $r_{ig}=0$ are non-outliers, so that the observations follow the original pseudo-model (\ref{model1}).
On the other hand, when $r_{ig}=1$, the pseudo-marginal distribution of $Y_{ig}$ has polynomial rate, namely, $p(Y_{ig} | \tau_g, \beta_g, \omega )\propto Y_{ig}^{-2\nu-k_{ig}}$ for $k_{ig}>0$, and particularly, the marginal distribution with $k_{ig}=1$ is $t$-distribution with $2\nu$ degrees of freedom. 
For the value of $\nu$, we simply set a small value (e.g. $\nu=1/2$ corresponding a Cauchy-like tail) instead of estimating it. 
The use of such two-component mixture distributions is appealing in terms of robustness as well efficiency of posterior inference \citep{hamura2022log}.

To stabilize estimation of $\tau_g$ through borrowing strength from information of the other groups, we introduce the following hierarchical prior distributions for $\tau_g$ and $\beta_g$.
\begin{equation}\label{model2}
\tau_g\sim N(m_\tau, \psi_\tau), \ \ \ \ \beta_{gk}\sim N(m_{\beta_k}, \psi_{\beta_k}),  \ \ \ k=1,\ldots,p, 
\end{equation}
independently for $g=1,\ldots,G$, where $m_\tau$ and $m_{\beta_k}$ are unknown mean parameters and $\psi_\tau$ and $\psi_{\beta_k}$ are unknown variance parameters. 
Then, we have the hierarchical model consisting of (\ref{model1}) and (\ref{model2}).
The unknown parameters in (\ref{model2}) are common among all the groups and can be precisely estimated by using all the samples, leading to stable estimation of subgroup-wise treatment effects $\tau_g$ even when the subgroup-wise sample size is small.
This is a common strategy in hierarchical Bayesian modeling, known as ``borrowing strength".

Instead of the normal prior for $\tau_g$ as given in (\ref{model2}), it would be useful to use the spike-and-slab (SS) prior \citep[e.g.][]{ishwaran2005spike}, given by 
\begin{equation}\label{model-SS}
\tau_g\sim \pi N(0, \ep \psi_\tau) + (1-\pi)N(m_\tau, \psi_\tau),
\end{equation}
where $\pi$ is the unknown proportion and $\ep$ is a small (fixed) parameter. 
Note that the first part represents the null part, so that the most probability mass is concentrated around the origin. 
One advantage of the SS prior is to shrink $\tau_g$ toward to when the treatment effect in the $g$th groups is almost zero.

\subsection{Posterior computation}
\label{sec:pos}

The unknown parameters in the proposed hierarchical models consisting of (\ref{model1}) and (\ref{model2}) are $m_\tau, m_{\beta_k}$, $\psi_\tau, \psi_{\beta_k}$, $\omega$ and $w$. 
We assign (conditionally) conjugate priors for these parameters, $m_\tau, m_{\beta_k}\sim N(a_m, b_m)$, $\psi_\tau, \psi_{\beta_k}\sim {\rm IG}(a_\psi, b_\psi)$, $w\sim {\rm Beta}(a_w, b_w)$ and $\omega\sim {\rm Ga}(a_{\omega}, b_{\omega})$, where ${\rm IG}(a, b)$ denotes an inverse-gamma distribution with shape parameter $a$ and rate parameter $b$. 
In our numerical studies, we set $a_m=0$, $b_m=10^3$, $a_\psi=b_\psi=a_\omega=b_\omega=1$ and $a_w=b_w=1/2$ as default settings of the prior distributions. 
Note that the prior of $\tau_g$ can be expressed as 
$$
\tau_g|(s_g=1)\sim N(0, \ep \psi_\tau), \ \ \ \tau_g|(s_g=0)\sim N(m_\tau,\psi_\tau), \ \ \ P(s_g=0)=1-P(s_g=1)=\pi,
$$
independently for $g=1,\ldots,G$, where $s_g$ is a latent variable and $\ep$ is a fixed small value. 
We assign the prior of $\pi$ as $\pi\sim {\rm Beta}(a_\pi, b_\tau)$.

The posterior computation of the proposed hierarchical model can be done by a simple Gibbs sampling, where the detailed steps are described as follows: 

\begin{itemize}
\item[-]
(Sampling from $\omega$) \ Generate $\omega$ from ${\rm Ga}(a_\omega+\sum_{g=1}^G\sum_{i=1}^{n_g}k_{ig}/2,b_\omega+\sum_{g=1}^G\sum_{i=1}^{n_g}u_{ig}k_{ig}(Y_{ig}-\tau_gW_{ig}-Z_{ig}^\top\beta_g)^2/2)$.

\item[-]
(Sampling from $\tau_g$) \ Generate $\tau_g$ from $N(A_{\tau}B_{\tau}, A_{\tau})$, where 
\begin{align*}
A_\tau&=\left(\omega\sum_{i=1}^{n_g}u_{ig}k_{ig}W_{ig}+\frac{1}{\ep^{s_g}\psi_\tau}\right)^{-1}, \\ B_{\tau}&=\omega\sum_{i=1}^{n_g}u_{ig}k_{ig}W_{ig}(Y_{ig}-Z_{ig}^\top\beta_g)+(1-s_g)\frac{m_\tau}{\psi_\tau}.
\end{align*}

% for presentation
%\begin{align*}
%\left(\omega\sum_{i=1}^{n_g}k_{ig}W_{ig}+\frac{1}{\psi_\tau}\right)^{-1}\left\{\omega\sum_{i=1}^{n_g}k_{ig}W_{ig}(Y_{ig}-Z_{ig}^\top\beta_g)+\frac{m_\tau}{\psi_\tau}\right\}.
%\end{align*}

\item[-]
(Sampling from $s_g$) \ 
Generate $s_g$ from a Bernoulli distribution with success probability being $\{1+\pi L_1/(1-\pi)L_0\}$, where $L_1=\phi(\tau_g; 0, \ep\phi_\tau)$ and $L_0=\phi(\tau_g; m_\tau, \psi_\tau)$.

\item[-]
(Sampling from $\pi$) \ 
For $g=1,\ldots,G$, generate $\pi$ from ${\rm Beta}(a_{\pi}+\sum_{g=1}^Gs_g, b_{\pi}+G-\sum_{g=1}^Gs_g)$.

\item[-]
(Sampling from $\beta_g$) \  Generate $\beta_g$ from $N(A_{\beta}B_{\beta}, A_{\beta})$ independently for $g=1,\ldots,G$, where 
$$
A_{\beta}=\left(\omega\sum_{i=1}^{n_g}u_{ig}k_{ig}Z_{ig}Z_{ig}^\top+D_{\beta}^{-1}\right)^{-1}, \ \ \ \ B_{\beta}=\omega\sum_{i=1}^{n_g}u_{ig}k_{ig}Z_{ig}(Y_{ig}-\tau_gW_{ig})+D_{\beta}^{-1}\mu,
$$
with $D_{\beta}={\rm diag}(\delta_1,\ldots,\delta_p)$ and $\mu=(\mu_1,\ldots,\mu_p)$.

\item[-]
(Sampling from $\psi_\tau$) \ 
Generate $\psi_\tau$ from ${\rm IG}(a_{\psi}+G/2, b_{\psi}+\sum_{g=1}^G\ep^{-s_g}(\tau_g-m_{\tau})^2/2)$.

\item[-]
(Sampling from $\psi_{\beta_k}$) \ 
Generate $\psi_{\beta_k}$ from ${\rm IG}(a_{\psi}+G/2, b_{\psi}+\sum_{g=1}^G(\beta_{gk}-m_{\beta_k})^2/2)$ independently for $k=1,\ldots,p$.

\item[-]
(Sampling from $m_{\tau}$) \ 
Generate $m_{\tau}$ from $N(A_{m_\tau}B_{m_\tau}, A_{m_\tau})$, where
$$
A_{m_\tau}=\left\{\frac{\sum_{g=1}^G(1-s_g)}{\psi_\tau}+\frac{1}{b_m}\right\}^{-1}, \ \ \ B_{m_\tau}=\frac{1}{\psi_\tau}\sum_{g=1}^G (1-s_g)\tau_g + \frac{a_m}{b_m}.
$$

\item[-]
(Sampling from $m_{\beta_k}$) \ 
Generate $m_{\beta_k}$ from $N(A_{m_{\beta_k}}B_{m_{\beta_k}}, A_{m_{\beta_k}})$ independently for $k=1,\ldots,p$, where
$$
A_{m_{\beta_k}}=\left(\frac{G}{\psi_{\beta_k}}+\frac{1}{b_m}\right)^{-1}, \ \ \ B_{m_{\beta_k}}=\frac{1}{\psi_{\beta_k}}\sum_{g=1}^G \tau_g + \frac{a_m}{b_m}.
$$

\item[-]
(Sampling from $u_{ig}$) \  
Given $r_{ig}=0$, generate $u_{ig}$ from $\delta_1$ (i.e. setting $u_{ig}=1$) and given $r_{ig}=1$, generate $u_{ig}$ from
$$
{\rm Ga}\left(\nu+\frac{k_{ig}}{2}, \nu + \frac12\omega k_{ig}(Y_{ig}-\tau_gW_{ig}-Z_{ig}^\top \beta_g)^2\right),
$$
independently for $i=1,\ldots,n_g$ and $g=1,\ldots,G$.

\item[-]
(Sampling from $r_{ig}$) \ 
Generate $r_{ig}$ from a Bernoulli distribution with success probability being $1/\{1+(1-\pi)L_{0ig}/\pi L_{1ig}\}$ independently for $i=1,\ldots,n_g$ and $g=1,\ldots,G$, where $L_{1ig}=\phi(Y_{ig}; \tau_gW_{ig}+Z_{ig}^\top \beta_g,  \omega^{-1}u_{ig}^{-1})^{k_{ig}}$, $L_{0ig}=\phi(Y_{ig}; \tau_gW_{ig}+Z_{ig}^\top \beta_g,  \omega^{-1})^{k_{ig}}$.

\item[-]
(Sampling from $w$) \
Generate $w$ from ${\rm Beta}(a_{w}+\sum_{g=1}^G\sum_{i=1}^{n_g}r_{ig}, b_{w}+N-\sum_{g=1}^G\sum_{i=1}^{n_g}r_{ig})$.

\end{itemize}

\subsection{HRDD with binary response}\label{sec:bin}
The proposed HRDD is also applicable to other types of responses by changing the model and corresponding loss function. 
Here we provide HRDD under binary responses, namely, $Y_{ig}\in \{0, 1\}$.
We first assume that the conditional probability of $Y_{ig}$ is expressed as the following logistic model: 
$$
P(Y_{ig}=1|\tau_g, \beta_g)=
\frac{\exp(\mu_{ig}Y_{ig})}{1+\exp(\mu_{ig})}, \ \ \ \ 
\mu_{ig} =\tau_gW_{ig}+Z_{ig}^\top\beta_g.
$$
Note that \cite{xu2017regression} also adopted logistic models for RDD with categorical responses. 
Under the logistic model, the treatment effect can be defined as 
\begin{align}
\tau_g^{\ast}
&\equiv \lim_{x\to c+0}E[Y_{ig}|X_{ig}=x]-\lim_{x\to c-0}E[Y_{ig}|X_{ig}=x] \notag\\
&={\rm logistic}(\tau_g+\beta_{g1}) - {\rm logistic}(\beta_{g1}), \label{binary-tau}
\end{align}
where ${\rm logistic}(x)=\{1+\exp(-x)\}^{-1}$.
To adopt the above model in HRDD, we use the pseudo-model based on a weighted log-likelihood for the first stage distribution.
The pseudo-model is given by
$$
p(Y_{ig}|\tau_g, \beta_g)\propto 
\frac{\exp(\mu_{ig}Y_{ig}k_{ig})}{\{1+\exp(\mu_{ig})\}^{k_{ig}}}, \ \ \ \ 
\mu_{ig} =\tau_gW_{ig}+Z_{ig}^\top\beta_g,
$$
which can be regarded as power likelihood of the original logistic model \citep{holmes2017assigning,bissiri2016general}.
Note that the pseudo-model reflects the influence of the local information around the boundary of RDD through $k_{ig}$ in the same way as the weighted quadratic loss (\ref{loss}).

We assign $\tau_g\sim N(m_{\tau}, \psi_{\tau})$ and $\beta_{gk}\sim N(m_{\beta_k}, \psi_{\beta_k})$ for $k=1,\ldots,p$, where $m_\tau, m_{\beta_k}\sim N(a_m, b_m)$ and $\psi_{\tau}, \psi_{\beta_k}\sim {\rm IG}(a_\psi, b_\psi)$.
To estimate the pseudo-model, we can develop an efficient sampling scheme by using the P\'olya-gamma data augmentation \citep{polson2013bayesian}, given by 
$$
\frac{\exp(\mu_{ig}Y_{ig}k_{ig})}{\{1+\exp(\mu_{ig})\}^{k_{ig}}}
=2^{-k_{ig}}\exp(\kappa_{ig}\mu_{ig})\int \exp(-\omega_{ig}\mu_{ig}^2/2) f_{\rm PG}(\omega_{ig}; k_{ig}, 0)d\omega_{ig},
$$
where $\kappa_{ig}=k_{ig}(Y_{ig}-1/2)$ and $f_{\rm PG}(\cdot; b, c)$ is the density of the P\'olya-gamma distribution with parameters $b$ and $c$.
Owing to the data augmentation, the posterior computation can be carried out by a simple Gibbs sampler, described as follows: 
\begin{itemize}
\item[-] 
(Sampling from $\omega_{ig}$) \ 
Generate $\omega_{ig}$ from ${\rm PG}(k_{ig}, \mu_{ig}^2)$ independently for $i=1,\ldots,n_g$ and $g=1,\ldots,G$.
We use the R package ``BayesLogit" to generate random samples of the P\'olya-gamma distribution. 

\item[-]
(Sampling from $\tau_g$) \ 
Generate $\tau_g$ from $N(A_{\tau}B_{\tau}, A_{\tau})$ independently for $g=1,\ldots,G$, where 
$$
A_\tau=\left(\sum_{i=1}^{n_g}\omega_{ig}W_{ig}+\frac{1}{\psi_\tau}\right)^{-1}, \ \ \ \ B_{\tau}=\sum_{i=1}^{n_g}W_{ig}(\kappa_{ig}-\omega_{ig}Z_{ig}^\top\beta_g)+\frac{m_{\tau}}{\psi_\tau}.
$$

\item[-]
(Sampling from $\beta_g$) \ 
Generate $\beta_g$ from $N(A_{\beta}B_{\beta}, A_{\beta})$ independently for $g=1,\ldots,G$, where 
$$
A_{\beta}=\left(\sum_{i=1}^{n_g}\omega_{ig}Z_{ig}Z_{ig}^\top+D_{\beta}^{-1}\right)^{-1}, \ \ \ \ B_{\beta}=\sum_{i=1}^{n_g}Z_{ig}(\kappa_{ig}-\omega_{ig}\tau_gW_{ig})+D_{\beta}^{-1}M,
$$
with $D_{\beta}={\rm diag}(\psi_{\beta_1},\ldots,\psi_{\beta_p})$ and $M=(m_{\beta_1},\ldots,m_{\beta_p})$.

\item[-]
(Sampling from $\psi_\tau$, $\psi_{\beta_k}$, $m_{\tau}$ and $m_{\beta_k}$) \ The full conditional distributions are same as those under continuous response.
\end{itemize}

The posterior samples of the treatment effect $\tau_g^{\ast}$ defined in (\ref{binary-tau}) can be obtained through the posterior samples of $\tau_g$ and $\beta_{g1}$.

\subsection{Adaptation of bandwidth}\label{sec:band}
To carry out RDD, it is crucial to set the subgroup-wise bandwidth parameter $h_g$, which would control bias and efficiency of the resulting estimator. 
To estimate $h_g$, we first consider the leave-one-out pseudo-predictive distribution for the out-of-sample observation based on the pseudo-model (\ref{model1}): 
$$
p_h(Y_{ig}|Y_{-i,g})=\int p_h(Y_{ig}|\theta_{ig})\pi_h(\theta_{ig}|Y_{-i,g})d\theta_{ig},
$$
where $Y_{-i,g}$ is the set of observations in the $g$th group expect for $Y_{ig}$, $\theta_{ig}\equiv (\tau_g, \beta_g, \omega, u_{ig})$, $p_h(Y_{ig}|\theta_{ig})$ defined in (\ref{model2}) and $\pi_h(\theta_{ig}|Y_{-i,g})$ is the marginal posterior distribution of $\theta_{ig}$ given the data $Y_{-i,g}$ and bandwidth $h$.
Since the pseudo-prediction distribution $p_h(Y_{ig}|Y_{-i,g})$ is not necessarily normalized, we evaluate the predictive distribution by the Hyv\"{a}rinen score \citep{hyvarinen2005estimation,shao2019bayesian} as a function of $h$. 
In particular, to carry our evaluation around the threshold value, we propose the following local Hyv\"{a}rinen score:
$$
H_g(h)=\sum_{i\in T_g^m} \left[2\frac{\partial^2}{\partial Y_{ig}^2}\log p_h(Y_{ig}|Y_{-i,g})+\left(\frac{\partial}{\partial Y_{ig}}\log p_h(Y_{ig}|Y_{-i,g})\right)^2\right],
$$
where $T_g^m$ is the evaluation set defined as the $m$-nearest observations to the threshold in the $g$th group, and we set $m$ to $\max(0.02 n_g, 5)$ in each group through our numerical studies. 
As noted in \cite{yonekura2023adaptation}, the above criterion can be expressed as 
\begin{equation}\label{H-score}
H_g(h)
=\sum_{i\in T_g^m} 
\bigg\{
2E\Big[\ell^{(2)}_h(\theta_{ig}) + \big\{\ell^{(1)}_h(\theta_{ig})\big\}^2\Big] 
-\Big(E\left[\ell^{(1)}_h(\theta_{ig})\right]\Big)^2\bigg\},
\end{equation}
where the expectation is taken with respect to the marginal posterior distribution of $\theta_{ig}$ and  
$$
\ell^{(k)}_h(\theta_{ig})\equiv \frac{\partial^{k}\log p_h(Y_{ig}|\theta_{ig})}{\partial^{k}Y_{ig}},\ \ \ \ \ k=1,2.
$$
From (\ref{model1}), it follows that 
$$
\ell^{(1)}_h(\theta_{ig})= -\frac{\omega k_{ig}}{u_{ig}}(Y_{ig}-\tau_gW_{ig}-Z_{ig}^\top \beta_g), \ \ \ \ \ 
\ell^{(2)}_h(\theta_{ig})= -\frac{\omega k_{ig}}{u_{ig}}.
$$
To cast the score (\ref{H-score}) for estimating $h_g$, we first prepare a set of candidate values $\{a_1,\ldots,a_L\}$.
Starting the MCMC algorithm with $h_g=a_1$ for all $g$, we compute the Hyv\"{a}rinen score (\ref{H-score}) for a small batch of MCMC iterations and set $h_g=a_2$ to compute the score based on next small batch of MCMC iterations. 
If the new score is smaller than the previous one, we change $h_g$ to the next value (e.g. $h_g=a_3$), but if the new score is larger than the previous one, we change $h_g$ to the previous value (e.g. $h_g=a_1$) and stop updating. 
Such procedures are repeated until all $h_g$ are convergent.
We call the algorithm ``local" selection.
On the other hand, assuming $h_g=h$ (i.e. common bandwidth for all the groups), we can select the optimal value of $h$ according to the average Hyv\"{a}rinen score, $G^{-1}\sum_{g=1}^G H_g(h)$, which we call ``global" selection.

Regarding the Hyv\"{a}rinen score under binary responses as discussed in Section~\ref{sec:bin}, we use the definition of the Hyv\"{a}rinen score for discrete outcomes given by \cite{matsubara2023generalized}.
Then, the local Hyv\"{a}rinen score can be defined as 
\begin{align*}
H_g(h)
&=\sum_{i\in T_g^m}\left[\left(\frac{ p_h(1-Y_{ig}|Y_{-i,g})}{ p_h(Y_{ig}|Y_{-i,g})}\right)^2 - \frac{2 p_h(Y_{ig}|Y_{-i,g})}{ p_h(1-Y_{ig}|Y_{-i,g})}\right].
\end{align*}
We note that 
\begin{align*}
\frac{ p_h(1-Y_{ig}|Y_{-i,g}) }{ p_h(Y_{ig}|Y_{-i,g})}
&=\int p_h(1-Y_{ig}|\theta_{ig}, Y_{-i,g})\frac{\pi(\theta_{ig}|Y_{-i,g})}{p_h(Y_{ig}|Y_{-i,g})}d\theta_{ig}\\
&=\int \frac{p_h(1-Y_{ig}|\theta_{ig}, Y_{-i,g})}{p_h(Y_{ig}|\theta_{ig}, Y_{-i,g})}\pi(\theta_{ig}|Y_{-i,g}, Y_{ig})d\theta_{ig}\\
&=\int \frac{p_h(1-Y_{ig}|\theta_{ig})}{p_h(Y_{ig}|\theta_{ig})}\pi(\theta_{ig}|Y_g)d\theta_{ig},
\end{align*}
so that the local Hyv\"{a}rinen score can be expressed as the posterior expectation as follows: 
\begin{equation}\label{H-score-bin}
H_g(h)
=\sum_{i\in T_g^m}\left\{\Big(E\left[\ell_h(\theta_{ig})\right]\Big)^2 - 2\Big(E\left[\ell_h(\theta_{ig})\right]\Big)^{-1}\right\},
\end{equation}
where $\ell_h(\theta_{ig})\equiv p_h(1-Y_{ig}|\theta_{ig})/p_h(Y_{ig}|\theta_{ig}) =\exp\big\{\mu_{ig}k_{ig}(1-2Y_{ig})\big\}$.
Then, we can adopt the same approach under continuous response to optimize $h_g$ within MCMC iterations.

%%-----------------------------------------------%%
%%                 Simulation                    %%
%%-----------------------------------------------%%
\section{Simulation study}
\label{sec:sim}

\subsection{Continuous response}\label{sec:sim-cont}

We demonstrate the usefulness of the proposed hierarchical RDD (HRDD), compared with a method that separately applies RDD to each group, denoted by separate RDD.
To this end, we set $G=100$ and equally divided $G$ groups into four clusters and set the same number of group-specific sample sizes $n_g$ to the same values within the same clusters. 
The pattern of the cluster-wise sample sizes is $(100, 200, 300, 400)$.
Then, the running variable $X_{ig} \ (i=1,\ldots,n_g, \ g=1,\ldots,G)$  is generated as $X_{ig}=2T_{ig}-1$ with $T_{ig}\sim {\rm Beta}(2,4)$.
We consider the three scenarios for the true subgroup-wise treatment effects: 
\begin{align*}
{\rm (I)} \ \  &\tau_g \sim {\rm Ga}(3, 1) -3,\\
{\rm (II)} \ \  &\tau_g\sim 0.4\delta_{-2} + 0.2\delta_0 + 0.4\delta_{2}, \\
{\rm (III)} \ \  &\tau_g\sim 0.4U(-3,-1) + 0.2\delta_0 + 0.4U(1,3), 
\end{align*}
where $\delta_a$ is the Dirac measure on $a$. 
We generate the response variable $Y_{ig}$ as 
$$
Y_{ig}=\mu_{ig}^{(1)}I(X_{ig}\leq c)+\mu_{ig}^{(2)}I(X_{ig}>c)+\sigma_g\ep_{ig}
$$
where 
\begin{align*}
&\mu_{ig}^{(1)}=\beta_{g1}^{(1)} X_{ig}+ \beta_{g2}^{(1)} X_{ig}^2 + \beta_{g3}^{(1)} X_{ig}^3, \\
&\mu_{ig}^{(2)}=\tau_g + \beta_{g1}^{(2)} X_{ig}+ \beta_{g2}^{(2)} X_{ig}^2 + \beta_{g3}^{(2)} X_{ig}^3.
\end{align*}
Here we set $c=0$, generated $\sigma_g^2$ from $U(0.5, 1.2)$ and $\ep_{ig}$ is an error term.
Regarding the subgroup-wise regression coefficients, we set $\beta_{g1}^{(1)}, \beta_{g1}^{(2)}\sim  U(0.4, 1.4)$,  $\beta_{g2}^{(1)}\sim U(3,7)$, $\beta_{g2}^{(2)}\sim U(5,9)$, $\beta_{g3}^{(1)}\sim U(9, 11)$ and $\beta_{g3}^{(2)}\sim U(3,5)$.
For the error term $\ep_{ig}$, we consider the following three scenarios:
\begin{align*}
{\rm (A)} \  \ep_{ig}\sim N(0,1), \ \ \ \ {\rm (B)} \ \ep_{ig}\sim t_3, \ \ \ \
{\rm (C)} \  \ep_{ig}\sim {\rm Ga}(4,2)-2, 
\end{align*}
where $t_3$ is the $t$-distribution with $3$ degrees of freedom. 
Note that some outlying observations are generated under Scenario (B), and the error distribution under Scenario (C) is skewed.

For the simulated dataset, we fit the proposed robust HRDD (with hierarchy for $u_{ig}$), where the bandwidth is selected by applying "local" method optimizing the group-wise bandwidth parameters by $H_g(h)$ in (\ref{H-score}), and "global" method optimizing the bandwidth parameter common to all the groups by $G^{-1}\sum_{g=1}^GH_g(h)$.
These methods are denoted by HRDD-L (for local method) and HRDD-G (for global method).
The triangular kernel is used in HRDD and normally distributed random effects are used for subgroup-wise treatment effects.  
We generated 1000 posterior samples after discarding the first 500 samples, and obtained posterior means and $95\%$ credible intervals of $\tau_g$. 
For comparison, we applied the ``rdrobust" package \citep{calonico2015rdrobust} separately to each group.
There are three options in the ``rdrobust" package, namely conventional (sRDD-c), bias-corrected (sRDD-bc), and robust (sRDD-r) methods \citep{calonico2014robust}.
Note that we used a local linear regression for both HRDD and sRDD and quadratic regression for bias correction in sRDD. 
We also applied the standard RDD assuming homogeneity for the group-wise treatment effects, using the ``rdrobust" package.
Further, we adapt the nonparametric Bayesian RDD \citep{chib2023nonparametric}, where the unknown functions of the running variable in treatment and control groups are modeled by the adaptive cubic spline \citep{chib2010additive}.
We adapted 7 and 15 knots for the spline functions, denoted by BRDD1 and BRDD2, respectively. 
We apply the Bayesian model separately to each group by generating 1000 posterior samples after the first 500 samples.

\begin{figure}[!htb]
\begin{center}
\includegraphics[width=16cm]{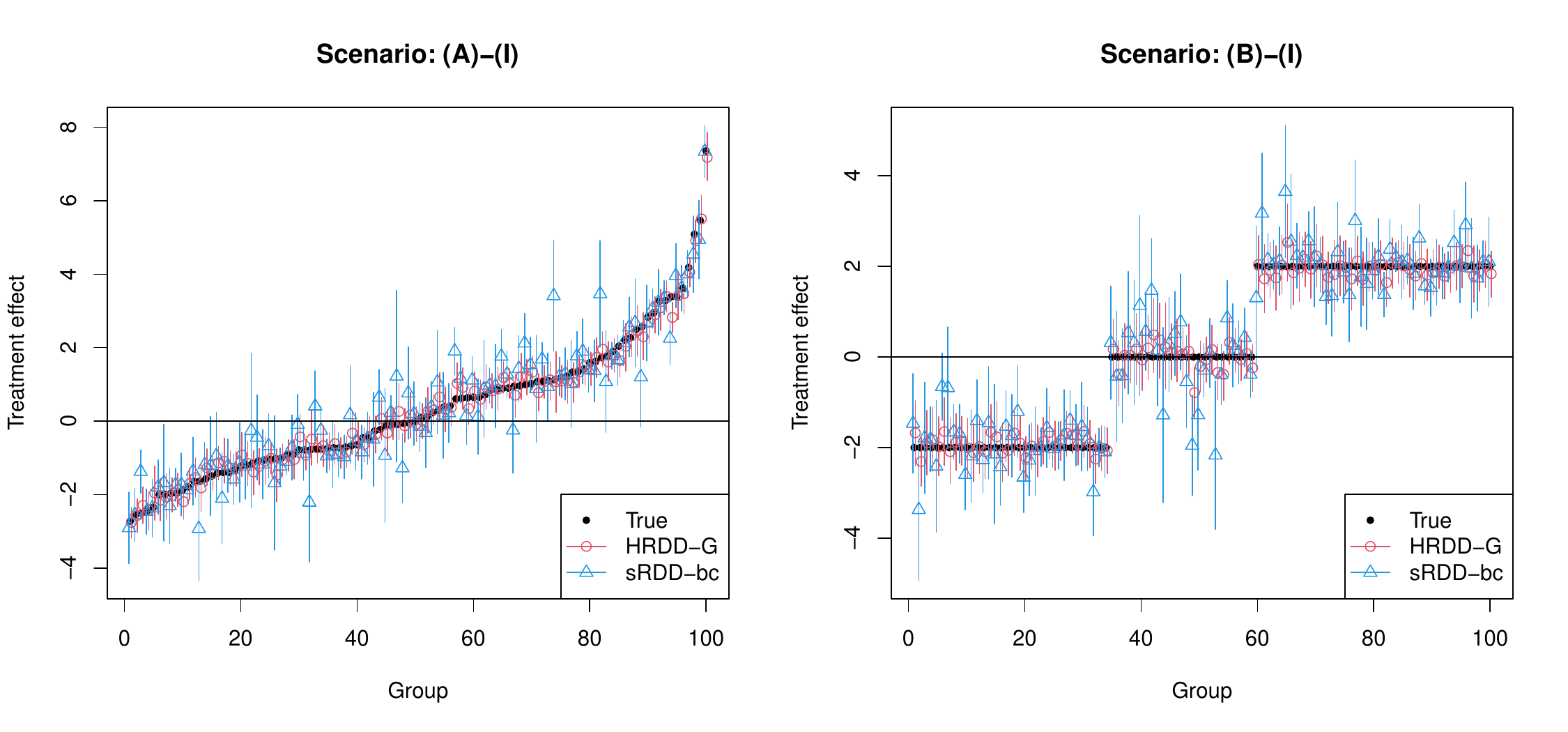}
\end{center}
\caption{Point estimates and $95\%$ confidence (credible) intervals of $\tau_g$ obtained by the proposed hierarchical RDD (HRDD-G) and separately applying the conventional RDD to each subgroup (sRDD-r) under four scenarios. }
\label{fig:oneshot}
\end{figure}

We first show the results for one simulated dataset. 
In Figure~\ref{fig:oneshot}, we present point estimates and $95\%$ confidence (credible) intervals of $\tau_g$ based on sRDD-r and HRDD-G. 
It shows that the point estimates of sRDD are highly variable and inaccurate while the proposed HRDD provides stable and accurate point estimates. 
Moreover, it can be seen that the lengths of confidence intervals by sRDD are much longer than those of credible intervals by HRDD, demonstrating the efficiency of HRDD owing to the effect of ``borrowing strength".

%  Figure 

To quantify the performance of the methods, we evaluated mean squared errors of point estimates and coverage performance of $95\%$ confidence (credible) intervals of $\tau_g$, based on 200 Monte Carlo replications, which are reported in Table~\ref{tab:sim-Gauss}.
It is observed that BRDD (a Bayesian version of RDD) tends to produce smaller RMSE than sRDD (a frequentist approach), and the proposed HRDD significantly outperforms two separate approaches (sRDD and BRDD). 
Regarding the interval estimation, the CP of all three types of sRDD tends to be smaller than the nominal level, while BRDD and HRDD attain coverage probability larger than the nominal level. 
Further, in spite of large CPs of HRDD, ALs of HRDD are much shorter than those of sRDD and BRDD, indicating the efficiency of interval estimation of HRDD owing to the benefit of borrowing strength caused by the hierarchical structure. 
Comparing the two types of HRDD, it can be seen that the use of global bandwidth selection (HRDD-G) tends to provide better estimates and more efficient credible intervals than HRDD-L.

To investigate the effect of group size (group-wise sample size), we consider the situation where all the groups have the same sample size (i.e., $n_g=n$) and evaluated RMSE under the scenario (B)-(II) with nine values of $n_g$.
Table~\ref{tab:sim-sample} reports the results based on 200 Monte Carlo replications.
It can be seen that RMSE decreases with the group size for all the methods, as expected. 
Moreover, the difference in RMSE between sRDD and BRDD gets smaller as the group size increases.
On the other hand, HRDD significantly improves RMSE over the other methods, even under large group sizes.

%  Table 
\begin{table}[thb!]
\caption{Root mean squared errors (RMSE) of point estimates, coverage probability (CP) and average length (AL) of $95\%$ confidence (credible) intervals of $\tau_g$ under 9 scenarios with continuous responses, based on separate RDD (sRDD), global (homogeneous) RDD, separate Bayesian RDD with 7 knots (BRDD1) and 15 knots (BRDD2) , and the proposed hierarchical RDD (HRDD) with local (-L) and global (-G) adaptation of the optimal bandwidth. 
}
\label{tab:sim-Gauss}
\begin{center}
Root Mean Squared Error of Point Estimates

{\footnotesize
\vspace{0.5mm}
\begin{tabular}{ccccccccccccccccccc}
\hline
Error& & \multicolumn{3}{c}{(A)} && \multicolumn{3}{c}{(B)} && \multicolumn{3}{c}{(C)} \\ 
$\tau$ &  & (I) & (II) & (III) &  & (I) & (II) & (III) &  & (I) & (II) & (III) \\
\hline
sRDD &  & 0.54 & 0.55 & 0.54 &  & 0.88 & 0.91 & 0.89 &  & 0.54 & 0.55 & 0.54 \\
sRDD-bc &  & 0.63 & 0.64 & 0.62 &  & 1.03 & 1.06 & 1.04 &  & 0.62 & 0.64 & 0.63 \\
RDD &  & 1.74 & 1.79 & 1.85 &  & 1.73 & 1.78 & 1.85 &  & 1.73 & 1.78 & 1.86 \\
BRDD1 &  & 0.45 & 0.45 & 0.44 &  & 0.47 & 0.47 & 0.47 &  & 0.43 & 0.43 & 0.43 \\
BRDD2 &  & 0.42 & 0.42 & 0.41 &  & 0.45 & 0.45 & 0.45 &  & 0.40 & 0.41 & 0.40 \\
HRDD-L &  & 0.26 & 0.27 & 0.26 &  & 0.32 & 0.33 & 0.32 &  & 0.26 & 0.26 & 0.25 \\
HRDD-G &  & 0.22 & 0.22 & 0.21 &  & 0.27 & 0.27 & 0.26 &  & 0.21 & 0.21 & 0.21 \\
\hline
\end{tabular}
}

\medskip
Coverage Probability (\%) of Credible Intervals

{\footnotesize
\vspace{0.5mm}
\begin{tabular}{cccccccccccccccc}
\hline
Error& & \multicolumn{3}{c}{(A)} && \multicolumn{3}{c}{(B)} && \multicolumn{3}{c}{(C)} \\ 
$\tau$ &  & (I) & (II) & (III) &  & (I) & (II) & (III) &  & (I) & (II) & (III) \\
\hline
sRDD &  & 91.4 & 91.4 & 91.5 &  & 92.6 & 92.8 & 92.3 &  & 91.9 & 91.9 & 91.7 \\
sRDD-bc &  & 86.8 & 87.1 & 86.8 &  & 88.0 & 88.2 & 87.9 &  & 87.7 & 87.6 & 87.5 \\
sRDD-r &  & 91.7 & 91.8 & 91.9 &  & 93.0 & 93.0 & 92.7 &  & 92.4 & 92.3 & 92.0 \\
RDD-r &  & 7.7 & 11.0 & 12.3 &  & 10.0 & 14.0 & 13.8 &  & 7.8 & 11.8 & 11.5 \\
BRDD1 &  & 99.1 & 98.9 & 99.0 &  & 99.3 & 99.4 & 99.3 &  & 99.2 & 98.9 & 99.1 \\
BRDD2 &  & 99.2 & 99.0 & 99.1 &  & 99.3 & 99.3 & 99.3 &  & 99.3 & 99.1 & 99.2 \\
HRDD-L &  & 97.9 & 97.7 & 98.0 &  & 98.7 & 98.6 & 98.7 &  & 98.1 & 98.2 & 98.3 \\
HRDD-G &  & 98.2 & 98.1 & 98.3 &  & 98.8 & 98.8 & 98.9 &  & 98.5 & 98.5 & 98.5 \\
\hline
\end{tabular}
}

\medskip
Average Length of Credible Intervals

{\footnotesize
\vspace{0.5mm}
\begin{tabular}{cccccccccccccccc}
\hline
Error& & \multicolumn{3}{c}{(A)} && \multicolumn{3}{c}{(B)} && \multicolumn{3}{c}{(C)} \\ 
$\tau$ &  & (I) & (II) & (III) &  & (I) & (II) & (III) &  & (I) & (II) & (III) \\
\hline
sRDD &  & 1.78 & 1.78 & 1.78 &  & 2.81 & 2.84 & 2.80 &  & 1.76 & 1.77 & 1.76 \\
sRDD-bc &  & 1.78 & 1.78 & 1.78 &  & 2.81 & 2.84 & 2.80 &  & 1.76 & 1.77 & 1.76 \\
sRDD-r &  & 2.11 & 2.12 & 2.11 &  & 3.34 & 3.37 & 3.32 &  & 2.09 & 2.10 & 2.09 \\
RDD-r &  & 0.36 & 0.37 & 0.38 &  & 0.45 & 0.46 & 0.46 &  & 0.36 & 0.37 & 0.38 \\
BRDD1 &  & 2.33 & 2.34 & 2.33 &  & 2.59 & 2.60 & 2.60 &  & 2.28 & 2.29 & 2.28 \\
BRDD2 &  & 2.18 & 2.19 & 2.18 &  & 2.42 & 2.43 & 2.42 &  & 2.12 & 2.13 & 2.12 \\
HRDD-L &  & 1.21 & 1.20 & 1.20 &  & 1.56 & 1.58 & 1.57 &  & 1.20 & 1.20 & 1.19 \\
HRDD-G &  & 1.08 & 1.06 & 1.05 &  & 1.35 & 1.36 & 1.35 &  & 1.06 & 1.06 & 1.07 \\
\hline
\end{tabular}
}
\end{center}
\end{table}

%  Table (different sample size)
\begin{table}[thb!]
\caption{Root mean squared errors (RMSE) of point estimates of $\tau_g$ under scenario (B)-(II) with different within-group sample sizes, obtained by separate RDD (sRDD), separate Bayesian RDD with 7 knots (BRDD1) and 15 knots (BRDD2), and the proposed hierarchical RDD (HRDD) with local (-L) and global (-G) adaptation of the optimal bandwidth. 
}
\label{tab:sim-sample}
\begin{center}
{\footnotesize
\vspace{0.5mm}
\begin{tabular}{cccccccccccccccc}
\hline
group size ($n_g$) &  & 100 & 200 & 300 & 400 & 500 & 700 & 900 & 1200 & 1500 \\
\hline
sRDD &  & 2.16 & 1.55 & 0.95 & 0.77 & 0.68 & 0.55 & 0.48 & 0.41 & 0.36 \\
sRDD-bc &  & 2.53 & 1.81 & 1.11 & 0.90 & 0.79 & 0.64 & 0.56 & 0.48 & 0.42 \\
BRDD1 &  & 0.66 & 0.53 & 0.49 & 0.45 & 0.43 & 0.40 & 0.38 & 0.36 & 0.34 \\
BRDD2 &  & 0.66 & 0.52 & 0.46 & 0.42 & 0.39 & 0.37 & 0.35 & 0.33 & 0.31 \\
HRDD-L &  & 0.56 & 0.42 & 0.35 & 0.29 & 0.26 & 0.23 & 0.20 & 0.18 & 0.16 \\
HRDD-G &  & 0.50 & 0.36 & 0.29 & 0.25 & 0.23 & 0.19 & 0.17 & 0.15 & 0.14 \\
\hline
\end{tabular}
}
\end{center}
\end{table}

\subsection{Binary response}

We next compare the proposed HRDD with the conventional separate methods under binary response. 
Using the same 9 scenarios of data generation, we generate binary responses $Y_{ig}$ as $Y_{ig}=I(Y_{ig}^{\ast}\geq 0)$, where $Y_{ig}^{\ast}$ is the latent continuous variable following the same data generating process used in Section~\ref{sec:sim-cont}.
Note that the logistic model is misspecified in all the data generating processes. 
For the simulated dataset, we applied HRDD based on the logistic model as explained in Section~\ref{sec:bin}, and sRDD by using ``rdrobust" package.
As in Section~\ref{sec:sim-cont}, we considered two methods for optimal bandwidth for HRDD and three options for sRDD, used a local liner regression for both HRDD and sRDD and quadratic regression for bias correction in sRDD. 
For HRDD, we generated 1000 posterior samples after discarding the first 500 samples, and obtained posterior means and $95\%$ credible intervals of $\tau_g$.
As in Section~\ref{sec:sim-cont}, we computed RMSE, CP, and AL based on 200 Monte Carlo replications, which are reported in Table~\ref{tab:sim-Bin}.
Overall, the relative performance of sRDD and HRDD does not change, namely, both HRDD methods provide more efficient point and interval estimation than the three types of sRDD.

%  Table 
\begin{table}[thb!]
\caption{Root mean squared errors (RMSE) of point estimates, coverage probability (CP) and average length (AL) of $95\%$ confidence (credible) intervals of $\tau_g$ under 9 scenarios with binary response, based on separate RDD (sRDD), global (homogeneous) RDD and hierarchical RDD (HRDD) with local (-L) and global (-G) adaptation of the optimal bandwidth. 
}
\label{tab:sim-Bin}
\begin{center}
Root Mean Squared Error of Point Estimates

{\footnotesize
\vspace{0.5mm}
\begin{tabular}{cccccccccccccccc}
\hline
Error& & \multicolumn{3}{c}{(A)} && \multicolumn{3}{c}{(B)} && \multicolumn{3}{c}{(C)} \\ 
$\tau$ &  & (I) & (II) & (III) &  & (I) & (II) & (III) &  & (I) & (II) & (III) \\
\hline
sRDD &  & 0.25 & 0.23 & 0.23 &  & 0.25 & 0.23 & 0.23 &  & 0.25 & 0.23 & 0.24 \\
sRDD-bc &  & 0.29 & 0.26 & 0.27 &  & 0.29 & 0.27 & 0.27 &  & 0.29 & 0.27 & 0.27 \\
RDD &  & 0.36 & 0.42 & 0.41 &  & 0.36 & 0.43 & 0.41 &  & 0.36 & 0.42 & 0.41 \\
%RDD-bc &  & 0.36 & 0.43 & 0.41 &  & 0.36 & 0.43 & 0.41 &  & 0.36 & 0.43 & 0.41 \\
HRDD-L &  & 0.10 & 0.08 & 0.09 &  & 0.10 & 0.08 & 0.09 &  & 0.10 & 0.08 & 0.09 \\
HRDD-G &  & 0.09 & 0.07 & 0.08 &  & 0.09 & 0.07 & 0.08 &  & 0.09 & 0.07 & 0.08 \\

\hline
\end{tabular}
}

\medskip
Coverage Probability (\%) of Credible Intervals 

{\footnotesize
\vspace{0.5mm}
\begin{tabular}{cccccccccccccccc}
\hline
Error& & \multicolumn{3}{c}{(A)} && \multicolumn{3}{c}{(B)} && \multicolumn{3}{c}{(C)} \\ 
$\tau$ &  & (I) & (II) & (III) &  & (I) & (II) & (III) &  & (I) & (II) & (III) \\
\hline
sRDD &  & 89.0 & 89.4 & 89.0 &  & 88.8 & 89.2 & 89.2 &  & 88.8 & 89.3 & 89.3 \\
sRDD-bc &  & 85.0 & 85.4 & 85.1 &  & 85.0 & 85.2 & 85.4 &  & 85.0 & 85.0 & 85.4 \\
sRDD-r &  & 89.5 & 89.7 & 89.5 &  & 89.5 & 89.5 & 89.8 &  & 89.6 & 89.5 & 89.8 \\
%RDD &  & 5.9 & 13.4 & 14.0 &  & 6.0 & 13.5 & 13.4 &  & 6.1 & 12.3 & 11.9 \\
%RDD-bc &  & 5.8 & 13.0 & 13.1 &  & 6.0 & 13.3 & 12.9 &  & 6.1 & 12.2 & 11.6 \\
RDD-r &  & 6.7 & 14.0 & 14.9 &  & 7.0 & 14.5 & 14.7 &  & 7.1 & 13.5 & 13.3 \\
HRDD-L &  & 98.5 & 99.4 & 98.7 &  & 98.4 & 99.3 & 98.8 &  & 98.5 & 99.2 & 98.8 \\
HRDD-G &  & 98.5 & 99.3 & 98.5 &  & 98.4 & 99.3 & 98.8 &  & 98.4 & 99.3 & 98.5 \\
\hline
\end{tabular}
}

\medskip
Average Length of Credible Intervals 

{\footnotesize
\vspace{0.5mm}
\begin{tabular}{cccccccccccccccc}
\hline
Error& & \multicolumn{3}{c}{(A)} && \multicolumn{3}{c}{(B)} && \multicolumn{3}{c}{(C)} \\ 
$\tau$ &  & (I) & (II) & (III) &  & (I) & (II) & (III) &  & (I) & (II) & (III) \\
\hline
sRDD &  & 0.82 & 0.74 & 0.75 &  & 0.81 & 0.74 & 0.75 &  & 0.81 & 0.74 & 0.75 \\
sRDD-bc &  & 0.82 & 0.74 & 0.75 &  & 0.81 & 0.74 & 0.75 &  & 0.81 & 0.74 & 0.75 \\
sRDD-r &  & 0.97 & 0.88 & 0.90 &  & 0.96 & 0.87 & 0.90 &  & 0.97 & 0.88 & 0.90 \\
%RDD &  & 0.10 & 0.10 & 0.09 &  & 0.10 & 0.10 & 0.09 &  & 0.10 & 0.10 & 0.09 \\
%RDD-bc &  & 0.10 & 0.10 & 0.09 &  & 0.10 & 0.10 & 0.09 &  & 0.10 & 0.10 & 0.09 \\
RDD-r &  & 0.11 & 0.11 & 0.11 &  & 0.11 & 0.11 & 0.11 &  & 0.11 & 0.11 & 0.11 \\
HRDD-L &  & 0.46 & 0.38 & 0.40 &  & 0.46 & 0.39 & 0.40 &  & 0.46 & 0.39 & 0.40 \\
HRDD-G &  & 0.43 & 0.35 & 0.37 &  & 0.43 & 0.35 & 0.37 &  & 0.43 & 0.35 & 0.37 \\
\hline
\end{tabular}
}
\end{center}
\end{table}

%%-----------------------------------------------%%
%%                 Application                   %%
%%-----------------------------------------------%%
\section{Application: subgroup treatment effects in Colombian scholarship}
\label{sec:app}

We demonstrate HRDD in an empirical application of Colombian scholarship data \citep{londono2020upstream}.
From 2014 to 2018, the Colombian government operated a large-scale scholarship program called Ser Pilo Paga. The scholarship loan covers the full tuition cost of attending any four-year or five-year undergraduate program in any government-certified high-quality university in Colombia. 
The outcome of interest is enrollment (binary variable) in any college; hence, the policy impact may be heterogeneous by the characteristics of subjects.

We considered two categories of gender (``M" and ``F"), three categories of SES (``SES1", ``SES2" and ``SES3"), three categories of education levels of father (``fED1", ``fDC2" and ``fDC3") and mother (``mDC1", ``mDC2" and ``mDC3"), and two categories of schools (``Private" and ``Public"), which leads to 108 subgroups.
We then identified two subgroups in which the proportion of 
samples whose covariate is around the threshold is smaller than 0.1$\%$, which results in 106 subgroups and 274251 total sample size in the subsequent analysis.

\begin{figure}[!htb]
\centering
\includegraphics[width=14cm,clip]{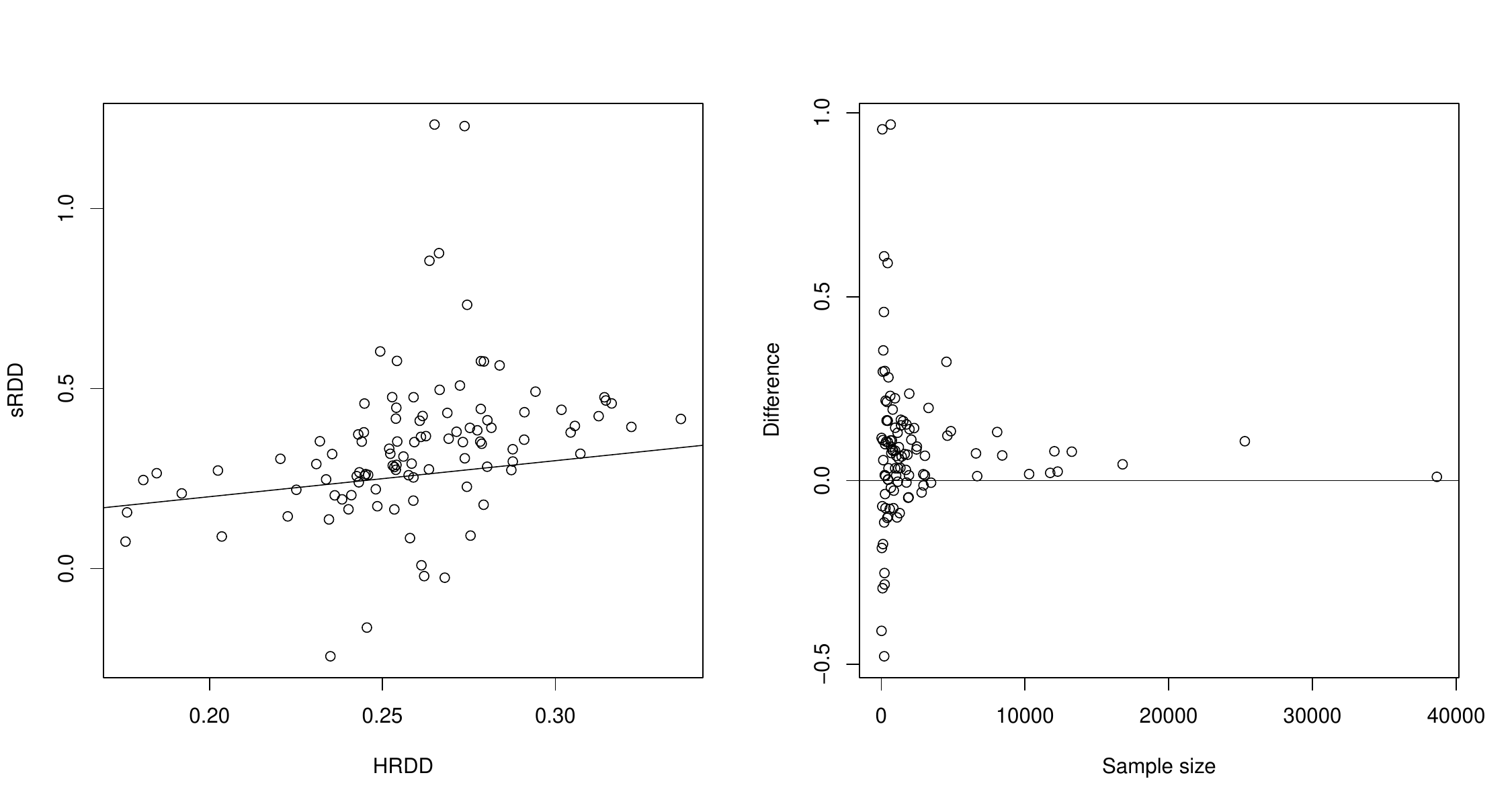}
\caption{Scatter plots of point estimates of HRDD-G and sRDD-c (left) and difference of these two estimates against subgroup-wise sample sizes (right).  }
\label{fig:app1}
\end{figure}

For the dataset, we applied both HRDD-G and HRDD-L with spike-and-slab priors on the group-wise treatment effects as well as sRDD-c and sRDD-bc (separately applying ``rdrobust" package to each subgroup). 
First, we found that sRDD methods produce numerical errors in 6 groups, possibly due to the small sample sizes around the thresholds.
Then, the subgroup-wise bandwidths used in HRDD methods in the 6 groups are replaced with the average value of the subgroup-wise bandwidths in other subgroups. 
Secondly, since the results of HRDD-G and HRDD-L are almost the same, we only show the results of HRDD-G in what follows. 
To compare HRDD and sRDD methods, we focus on 100 groups where the estimates of sRDD methods are obtained. 
In the left panel of Figure~\ref{fig:app1}, we present a scatter plot of the estimates of sRDD-c and HRDD, which shows that the estimates of sRDD-c are highly variable and are unrealistic values (more than 1) in some subgroups. 
Further, in the right panel of Figure~\ref{fig:app1}, we report the difference between two estimates by HRDD and sRDD-c against the subgroup-wise sample size. 
It is observed that there is a considerable difference in subgroups having small sample sizes, but the difference is vanishing as the sample size increases. 
This result indicates that the HRDD can successfully introduce shrinkage estimation of the treatment effects.
Combining the results of simulation studies in Section~\ref{sec:sim}, the results of HRDD would be more reliable than those of sRDD.

\begin{figure}[!htb]
\centering
\includegraphics[width=11cm,clip]{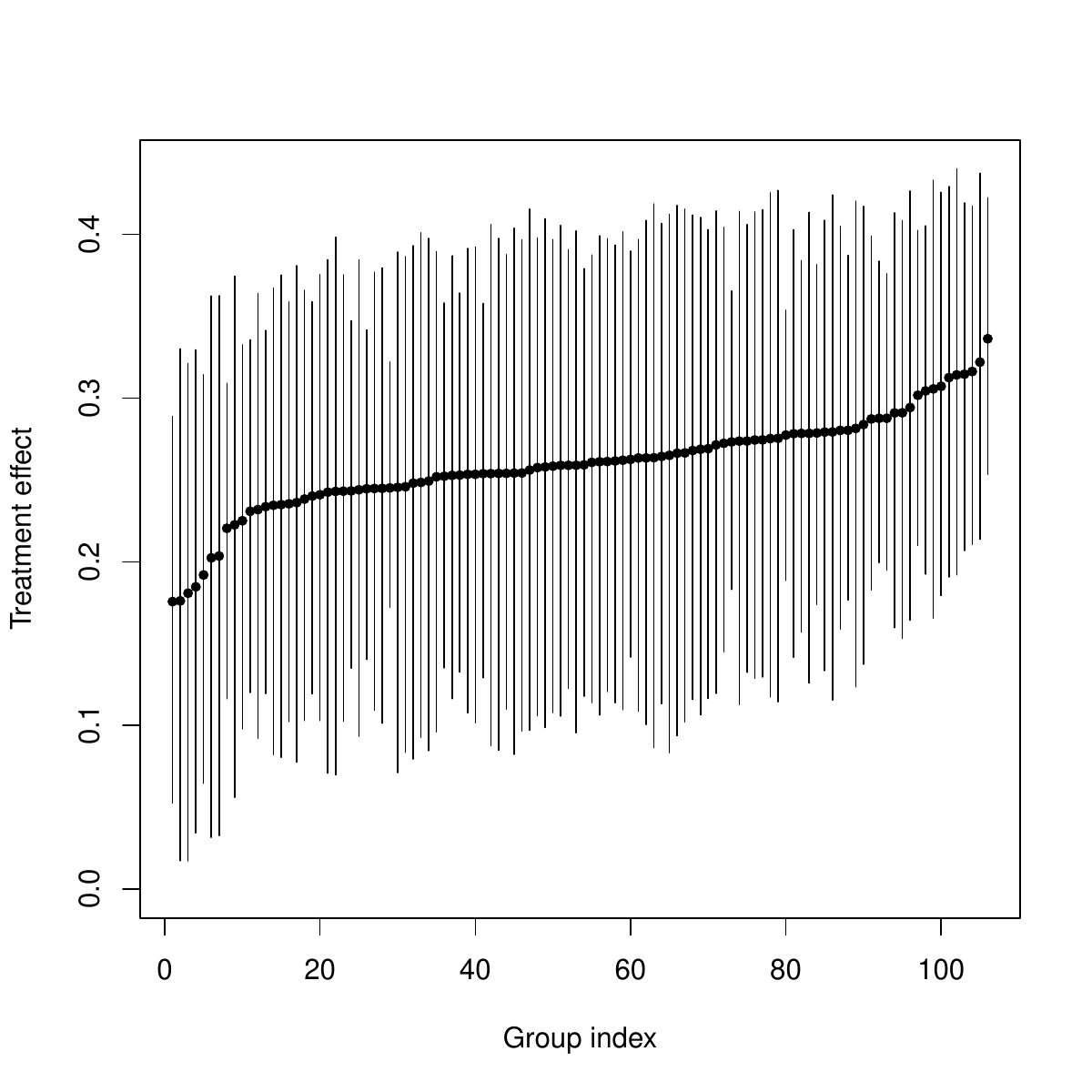}
\caption{Point estimates and $95\%$ confidence (credible) intervals of $\tau_g^{\ast}$ based on HRDD-G in the application to Colombian scholarship data. }
\label{fig:app2}
\end{figure}

Finally, in Figure~\ref{fig:app2}, we show the point estimates and $95\%$ credible intervals of the treatment effects for all the 106 groups, obtained by HRDD.
It can be seen that the estimates are around 0.25, which is relatively close to the average treatment effects reported in existing literature \citep[e.g.][]{sawada2024local} and has a certain heterogeneity among subgroups. 
The top three groups with the largest treatment effects are identified as (``M"$\times$``SES1"$\times$``fED2"$\times$``mED3"$\times$``Public"), (``M"$\times$``SES2"$\times$``fED3"$\times$``mED3"$\times$``Public") and (``F"$\times$``SES2"$\times$``fED2"$\times$``mED3"$\times$``Public").
Also, we can see that lengths of $95\%$ credible intervals are quite reasonable and all the credible intervals are bounded away from 0.

%%-----------------------------------------------%%
%%               Conclusion                      %%
%%-----------------------------------------------%%
\section{Concluding remarks}\label{sec:conc}
This paper introduces a hierarchical Bayesian framework for pursuing subgroup treatment effects under RDD.
Although the current paper is focused on sharp RDD for (subgroup-wise) average treatment effects with continuous and binary responses, our framework can be extended to various directions.
For example, by changing the loss functions (i.e. the pseudo-model for observations), our framework can also handle subgroup treatment effects under fuzzy RDD \citep{angrist1999using,van2002estimating}.
Furthermore, it is also possible to estimate other measures of treatment effects such as quantile treatment effects \citep[e.g.][]{frandsen2012quantile} by using check loss or equivalent adopting an asymmetric Laplace distribution as a pseudo-model. 
In this work, we considered using the H-score for the optimal bandwidth selection, which could be an example of the benefits of leveraging the Bayesian framework.
It would also be possible to consider model checking, comparison, and more flexible modeling of outcomes.
We left these potential extensions to valuable future works. 

While this paper employs a pseudo-model motivated by the local linear estimation, it would be possible to extend the nonparametric Bayesian RDD \citep{chib2023nonparametric} to handle subgroup treatment effects. 
However, the nonparametric Bayesian method uses all available data, including observations not close to the threshold, which may require considerable computational costs when introducing hierarchical structures. 
The detailed discussion will be left to a future study. 

While we employed the logistic function in the binary case presented in Section 2.3, other binary models (link functions) are available. 
A notable class of models is a latent variable model \citep{albert1993bayesian}, including the probit model and approximate logistic model using $t$-distribution as an error distribution for the latent variables. 
However, the power likelihood of these models does not give a simple Gibbs sampler, unlike using the standard likelihood. 
Hence, the logistic model combined with the P\'olya-gamma data augmentation would be preferable for efficient posterior computation.

\section*{Acknowledgement}
This work is supported by the Japan Society for the Promotion of Science (JSPS KAKENHI) grant numbers 21H00699 and 20H00080 (SUGASAWA), the JSPS KAKENHI grant number 23K12456 (KURISU), and the JSPS KAKENHI Grant Number 22K13373 (ISHIHARA).

\vspace{0.1cm}
%   Reference
\bibliographystyle{chicago}
\bibliography{ref}

\end{document}